\documentclass[final,1p,times,twocolumn]{elsarticle}

\journal{}

\usepackage{subfigure}
\usepackage{amsmath}
\usepackage{amssymb}
\usepackage{bm}
\usepackage{natbib}
\usepackage[colorlinks=true,citecolor=blue,urlcolor=blue,hyperfigures=true]{hyperref}
\usepackage{xcolor}  

\def\Eq{Eq.~}

\def\Fig{Fig.~}

\def\Ref{Ref.~}
\def\Refs{Refs.~}
\def\Sec{Sec.~}

\def\be{\begin{equation}}
\def\ee{\end{equation}}
\def\bea{\begin{eqnarray}}
\def\eea{\end{eqnarray}}

\def\ie{\textit{i.e.}~}

\newcommand{\ket}[1]{\left| #1 \right\rangle}

\newcommand{\refeqn}[1]{(\ref{#1})}

\begin{document}

\begin{frontmatter}




\title{The Sagnac effect: 20 years of development\\in matter-wave interferometry}


\author[LP2N]{B. Barrett}
\author[SYRTE]{R. Geiger}
\author[SYRTE]{I. Dutta}
\author[SYRTE]{M. Meunier}
\author[LP2N]{B. Canuel}
\author[LCAR]{A. Gauguet}
\author[LP2N]{P. Bouyer}
\author[SYRTE]{A. Landragin}

\address[LP2N]{LP2N, IOGS, CNRS and Universit\'{e} de Bordeaux, rue Fran\c{c}ois Mitterrand, 33400 Talence, France}
\address[SYRTE]{LNE-SYRTE, Observatoire de Paris, CNRS and UPMC, 61 avenue de l'Observatoire, 75014 Paris, France}
\address[LCAR]{Laboratoire Collisions Agr\'egats R\'eactivit\'e (LCAR), CNRS, Universit\'e Paul Sabatier, 118 route de Narbonne, 31062 Toulouse Cedex 09, France }

\begin{abstract}
Since the first atom interferometry experiments in 1991, measurements of rotation through the Sagnac effect in open-area atom interferometers has been studied. These studies have demonstrated very high sensitivity which can compete with state-of-the-art optical Sagnac interferometers. Since the early 2000s, these developments have been motivated by possible applications in inertial guidance and geophysics. Most matter-wave interferometers that have been investigated since then are based on two-photon Raman transitions for the manipulation of atomic wave packets. Results from the two most studied configurations, a space-domain interferometer with atomic beams and a time-domain interferometer with cold atoms, are presented and compared. Finally, the latest generation of cold atom interferometers and their preliminary results are presented.
\end{abstract}


\end{frontmatter}
%
%
\section{Introduction}
\label{sec:Introduction}

Rotation sensors are useful tools in both industry and fundamental scientific research. Highly accurate and precise rotation measurements are finding applications in inertial navigation \citep{Lawrence-Book-1998}, studies of geodesy and geophysics \citep{Igel-GeophysJInt-2007}, and tests of general relativity \citep{Will-LivingRevRel-2006}. Since the early 1900s, there have been many manifestations of Georges Sagnac's classic experiments \citep{Sagnac-CRAcadSci-1913} that utilize the ``Sagnac'' interference effect to measure rotational motion, both with light and with atoms \citep{Anderson-AJP-1994}. Gyroscopes based on this effect measure a rotation rate, $\Omega$, via a phase shift between the two paths of an interferometer. The Sagnac phase shift, for both photons and massive particles, can be written as:
\be
  \label{Phi_Sagnac}
  \Phi_{\rm Sagnac} = \frac{4\pi E }{h c^2} \bm{A} \cdot \bm{\Omega},
\ee
where $\bm{A}$ is the area vector of the Sagnac loop (normal to the plane of the interferometer and equal to the  area enclosed by the interferometer arms) and $E$ is the energy of the particle ($E = \hbar\omega$ for a photon of angular frequency $\omega$ and $E=Mc^2$ for a  particle of rest mass $M$). Equation \eqref{Phi_Sagnac} shows that the Sagnac phase for a matter wave interferometer is larger by a factor of $M c^2/\hbar \omega$ compared to an optical one with equivalent area. This scale factor is $\sim 10^{11}$ when comparing the rest energy of an atom to that of an optical photon in the visible range---emphasizing the high sensitivity of atom-based sensors to rotations. In this article, we will review some of the key developments that have taken place over the last 20 years regarding matter-wave Sagnac interferometers.

\begin{figure}[!t]
  \centering
  \includegraphics[width=0.9\textwidth]{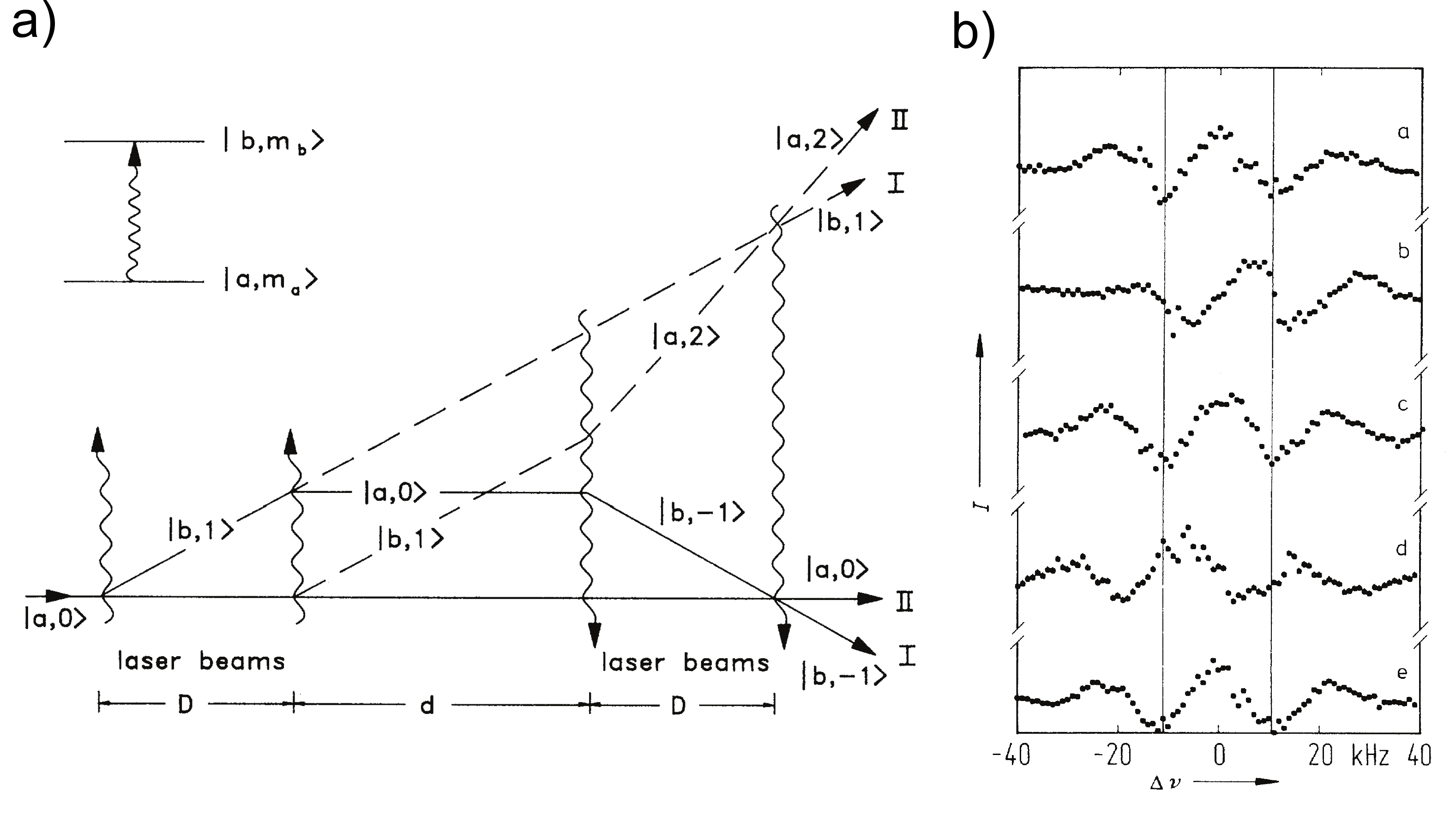}
  \caption{a) Ramsey-Bord\'{e} configuration of a state-labeled atom interferometer based on single-photon transitions. Here, a beam of atoms traverses two paires of traveling wave fields. The laser fields within each paire are separated by a distance $D$, while the two paires are separated by $d$ and are counter-propagating with respect to each other. b) Optical Ramsey fringes measured for the apparatus standing still (curves labeled a, c, and e), for the apparatus rotating at a rate of $\Omega = -90$ mrad/s (curve b), and for a rate $\Omega = +90$ mrad/s (curve d). The center of the Ramsey patterns for $\Omega = \pm 90$ mrad/s are clearly shifted to the right and left, respectively, relative to those for which $\Omega = 0$. Both figures were taken from \Ref \citep{Riehle-PRL-1991}.}
  \label{fig:Riehle}
\end{figure}

There has been dramatic progress in the field of atom interferometry in recent history. During the late 1980s, various types of atom interferometers were proposed as sensitive probes of different physical effects \citep{Borde-JPL-1983, Chebotayev-JOSAB-1985, Clauser-PhysicaB-1988, Borde-PhysLettA-1989}, and by the early 1990s the first experimental demonstrations had been realized \citep{Carnal-PRL-1991, Keith-PRL-1991, Kasevich-PRL-1991, Riehle-PRL-1991}. As a result of their intrinsically high sensitivity to inertial effects, atom interferometers are now routinely used as tools for studies of fundamental physics and precision measurements \citep{Cronin-RevModPhys-2009}. The first experiments that exploited the rotational sensitivity of atom interferometers were carried out by Riehle \textit{et al} \citep{Riehle-PRL-1991} using optical Ramsey spectroscopy with a calcium atomic beam. Figure \ref{fig:Riehle} shows their interferometer configuration and experimental results. By rotating their entire apparatus at various rates, $\Omega$, and recording the fringe shift of a Ramsey pattern, they were the first to demonstrate the validity of \Eq \refeqn{Phi_Sagnac} for atomic waves.

In 1997, two other research groups \citep{Lenef-PRL-1997, Gustavson-PRL-1997} simultaneously published results pertaining to rotation sensing with atom interferometers\footnote{In 1996, Oberthaler \textit{et al} \citep{Oberthaler-PRA-1996} also carried out sensitive rotation measurements with atoms using a Moir\'{e} deflectometer---a device consisting of an atomic beam and three mechanical gratings which can be considered the classical analog of a matter-wave interferometer. No quantum interference was involved in these measurements.}. Although both experiments relied on atomic beams, they each employed a different method to generate matter-wave interference.

\begin{figure}[!t]
  \centering
  \includegraphics[width=0.9\textwidth]{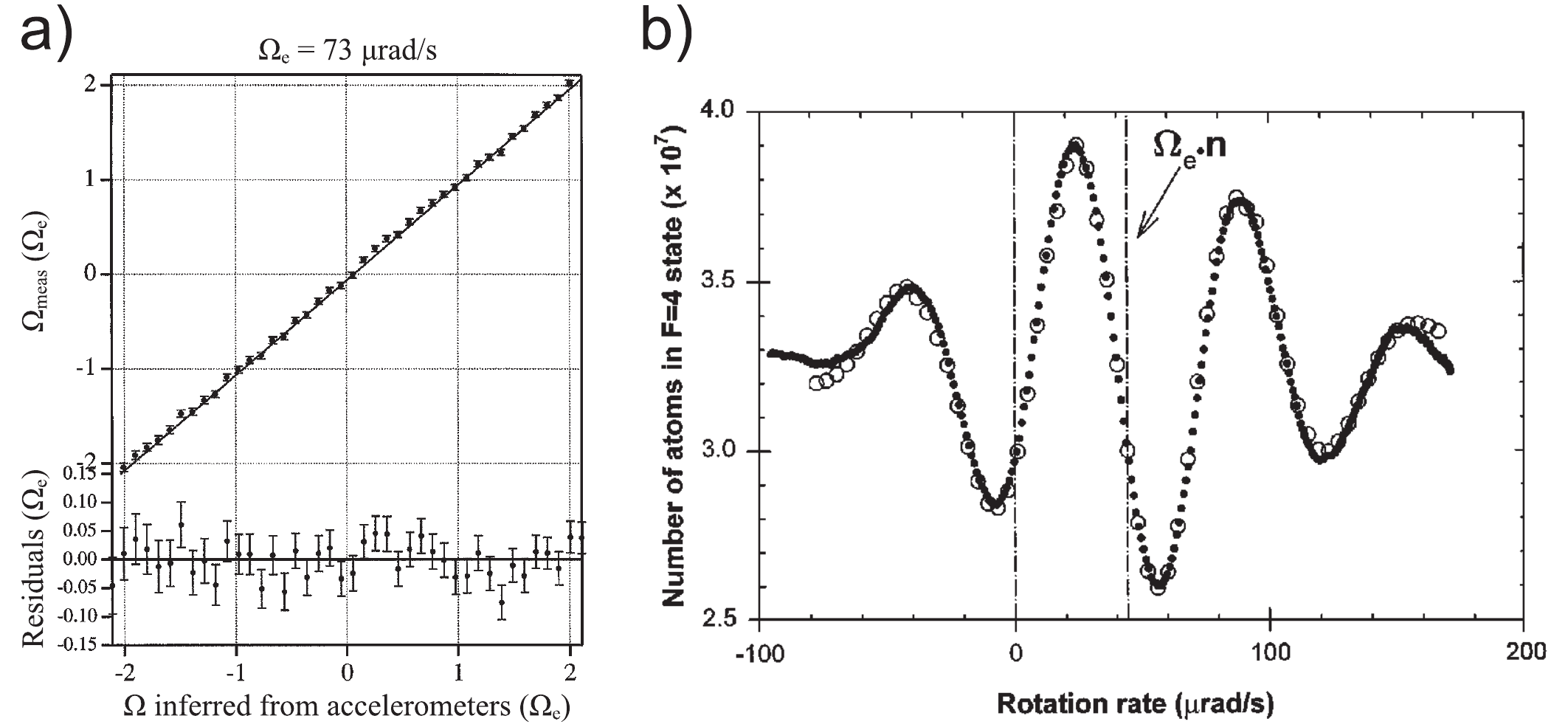}
  \caption{a) Experimental results from \Ref \citep{Lenef-PRL-1997}. Here, the rotation rate inferred from the interferometer, $\Omega_{\rm meas}$, is plotted with respect to the applied rate, $\Omega$, inferred from accelerometers attached to the apparatus. The slope of the linear fit was measured to be $1.008(7)$. The residuals of the fit are shown below. b) Atomic interference pattern as a function of applied rotation rate from \Ref \citep{Gustavson-PRL-1997}. The horizontal offset from zero rotation provides a direct measurement of the Earth's rotation rate, $\Omega_e$.}
  \label{fig:Lenef-Gustavson}
\end{figure}

In the work of \Ref \citep{Lenef-PRL-1997}, a beam of sodium atoms (longitudinal velocity $\sim 1030$ m/s) was sent through three nano-fabricated transmission gratings (200 nm period, 0.66 m separation) which acted to split, reflect and recombine atomic wavepackets taking part in the interferometer. By precisely controlling the applied rotation of their apparatus, they measured rotation rates of the same magnitude as that of the Earth ($\Omega_e = 73$ $\mu$rad/s), with a short-term sensitivity of about $3 \times 10^{-6}$ rad/s/$\sqrt{\rm Hz}$. Furthermore, they showed agreement with theory at the 1\% level over a relatively large range of $\pm 2 \Omega_e$---corresponding to a factor of 10 improvement over the first measurements of \Ref \citep{Riehle-PRL-1991}. Some of their experimental results are shown in \Fig \ref{fig:Lenef-Gustavson}a).

In contrast to \Ref \citep{Lenef-PRL-1997}, counter-propagating light pulses were used in \Ref \citep{Gustavson-PRL-1997} to manipulate a beam of cesium atoms (longitudinal velocity $\sim 290$ m/s). In this work, the atoms entered a $\sim 2$ m long interrogation region where they traversed three pairs of counter-propagating laser beams which drove a $\pi/2-\pi-\pi/2$ sequence of velocity-selective two-photon Raman transitions between long-lived hyperfine ground states. We explain in detail this interferometer scheme in \Sec \ref{subsec:principle}. Each pair of Raman beams was separated by 0.96 m and aligned perpendicular to the atomic trajectory. By rotating the Raman beams at different rates, an interference pattern was constructed in the number of $\ket{F = 4, m_F = 0}$ atoms at the output of the interferometer, as shown in \Fig \ref{fig:Lenef-Gustavson}b). The resulting short-term sensitivity of their rotation measurements was $2 \times 10^{-8}$ rad/s/$\sqrt{\rm Hz}$.

Comparing the short-term sensitivity achieved by these two experiments, there seems to be a clear advantage to using light pulses over nano-fabricated transmission gratings to split and recombine the atomic wavepackets (although some gain in sensitivity can be attributed the difference in the enclosed area between the two interferometers). The main advantage of using light pulses to interact with the atoms is their versatility and precision. One can easily modify the strength, bandwidth and phase of the light-matter interaction through precise control of the laser parameters. In comparison, nano-fabricated gratings are passive objects that must be carefully handled and placed within the vacuum system---making their modification or replacement much more challenging. For example, to change the phase of the gratings in \Ref \cite{Lenef-PRL-1997} by $\pi/2$ requires a physical displacement of only 50 nm perpendicular to the atomic trajectory. Modifying the phase of the light-matter interaction requires no moving parts, and can be done electro-optically with high precision. Furthermore, the use of two Raman lasers allows the use of state-labelling techniques to address the diffracted and undiffracted pathways of the interferometer \cite{Borde-PhysLettA-1989}. Usually, one detects the number of atoms remaining in either state by scattering many photons per atom, and inferring the phase shift from the ratio of state populations. This technique, which is not possible with transmission gratings, is less sensitive to fluctuations in total atom number and exhibits a high signal-to-noise ratio.

With the conclusion of these proof-of-principle experiments, the study of atomic gyroscopes entered a new phase which focused primarily on developing them as rotation sensors. This meant understanding and reducing sources of noise and systematic error, as well as improving the short-term sensitivity, linearity, long-term stability and accuracy of the devices. Furthermore, there remained a  question regarding the type of coherent matter-wave source to design the sensor around: atomic beams or cold atoms. Modern (since the year 2000) atomic rotation sensors and the improvement of their performances will be described below.

\label{par:outline}
The remainder of this article is organized as follows. In \Sec \ref{subsec:principle}, we review the basic operation principles of Sagnac interferometers based on two-photon Raman transitions, which represents the key experimental technique used in modern atomic gyroscopes. Section \ref{subsec:spacetime} presents two examples of experiments using respectively atomic beam and cold atoms, and the comparison of their performances. Section \ref{subsec:newgeneration} describes the most recent experiments of cold atom Sagnac interferometers. Finally, we give some perspectives for future improvements to these sensors and conclude in \Sec \ref{sec:conclusion}.

\section{Atomic rotation sensors}
\subsection{Principles of atomic Sagnac interferometers}
\label{subsec:principle}

In this section, we describe the basic operation principles of an atom-based gyroscope based on optical two-photon Raman transitions. All light-pulse interferometers work on the principle of momentum conservation between atoms and light. When an atom absorbs (emits) a photon of momentum $\hbar \bm{k}$, it undergoes a momentum impulse of $\hbar \bm{k}$ ($-\hbar \bm{k}$). In the case of Raman transitions, the momentum state of the atom is manipulated between two long-lived electronic ground states. Two laser beams with frequencies $\omega_1$ and $\omega_2$, respectively, are tuned such that their frequency difference, $\omega_1 - \omega_2$, is resonant with a microwave transition between two hyperfine ground states, which we label $\ket{1}$ and $\ket{2}$. When the Raman beams are counter-propagating (\ie when the wave vector $\bm{k}_2 \approx -\bm{k}_1$), a momentum exchange of approximately twice the single photon momentum accompanies these transitions: $\hbar (\bm{k}_1 - \bm{k}_2) \approx 2\hbar \bm{k}_1$. This results in a strong sensitivity to the Doppler frequency, $\bm{k}_{\rm eff} \cdot \bm{\varv}$, associated with the motion of the atoms, where $\bm{k}_{\rm eff} = \bm{k}_1 - \bm{k}_2$ is the effective $k$-vector of the light field. Under appropriate conditions, a Raman laser pulse can split the atom into a superposition of states $\ket{1,\bm{p}}$ and $\ket{2,\bm{p} + \hbar\bm{k}_{\rm eff}}$ (with a pulse area of $\pi/2$), or it can exchange these two states (with a pulse area of $\pi$). With these tools, it is possible to coherently split, reflect and recombine atomic wave packets such that they enclose a physical area---forming an interferometer that is sensitive to rotations.

\begin{figure}[!t]
  \centering
  \includegraphics[width=0.7\textwidth]{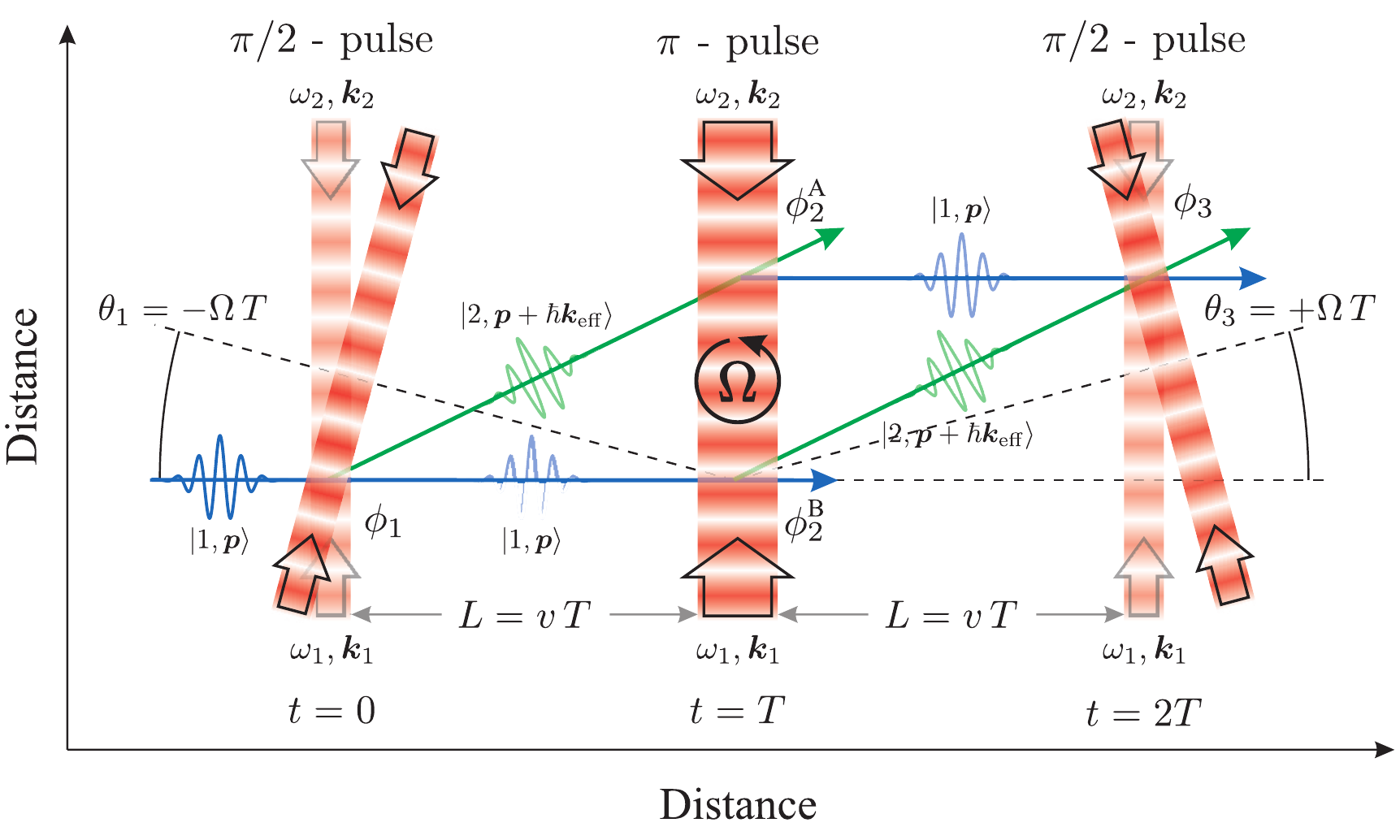}
  \caption{(Color online) Schematic of a matter-wave Sagnac interferometer based on two-photon Raman transitions. An atom in state $\ket{1}$, with center-of-mass velocity $\bm{\varv} = \bm{p}/M$, is subject to a sequence of counter-propagating laser pulses that are rotating relative to the atomic trajectory at a constant rate $\Omega$.}
  \label{fig:SagnacInterferometer}
\end{figure}

Figure \ref{fig:SagnacInterferometer} shows the most common matter-wave interferometer configuration, which consists of a $\pi/2-\pi-\pi/2$ sequence of Raman pulses, each separated by a time $T$ (analog to an optical Mach-Zehnder interferometer). If there is a phase shift between the wave packets associated with each internal state at the output of the interferometer, it manifests as a simple sinusoidal variation between the state populations:
\be
  \label{P2}
  \frac{N_2}{N_1 + N_2} = \frac{1 - \cos \Phi_{\rm tot}}{2}.
\ee
Here, $N_1$ and $N_2$ are the number of atoms in states $\ket{1,\bm{p}}$ and $\ket{2,\bm{p}+\hbar\bm{k}_{\rm eff}}$, respectively, and $\Phi_{\rm tot}$ is the total phase shift of the interferometer given by:
\be
  \label{Phi_tot}
  \Phi_{\rm tot} = (\phi_1 - \phi_2^{\rm A}) - (\phi_2^{\rm B} - \phi_3).
\ee
The individual phases, $\phi_i$, in this expression are imprinted on the atom by each Raman pulse. They take the form $\phi_i = \bm{k}_{\rm eff}^{(i)} \cdot \bm{r}(t_i) + \phi_L^{(i)}$, based on the orientation of the effective $k$-vector, $\bm{k}_{\rm eff}^{(i)}$, the position of the center of mass of the wave packet, $\bm{r}(t_i)$, and the relative phase between the two Raman lasers, $\phi_L^{(i)}$, at the time of the $i^{\rm th}$ pulse, $t = t_i$. The superscripts ``A'' and ``B'' on $\phi_2$ indicate the upper and lower pathways of the interferometer, respectively, as shown in \Fig \ref{fig:SagnacInterferometer}.

In general, there are two types of interferometer signals that can be detected within the realm of inertial effects: changes in absolute velocity (\ie accelerations) and changes in the velocity vector (\ie rotations). For accelerations, the sensitivity axis of the interferometer is along the propagation axis of the Raman lasers, while for rotations the interferometer is sensitive along an axis perpendicular to the plane defining the enclosed area. The evaluation of interferometer phase shifts in a non-inertial reference frame (accelerating or rotating) has been described in detail in previous publications \citep{Clauser-PhysicaB-1988, Storey-JPhysII-1994, Borde-CRAcadSci-2001, Borde-Metrologia-2002, Antoine-PhysLettA-2003}. Here, we give an intuitive calculation of the phase shift for an atom interferometer in a frame rotating at a constant rate. Figure \ref{fig:SagnacInterferometer} illustrates the situation from the atom's perspective, where the Raman lasers are rotating at a rate $\Omega$. At $t = 0$, the orientation of the Raman beams is rotated by an angle $\theta_1 = -\Omega T$ relative to the propagation axis of the atomic trajectory. Provided that $|\theta_1| \ll 1$, this imprints a phase shift on the atoms of $\phi_1 = k_{\rm eff} \theta_1 L$. At $t = T$, the Raman beam is perpendicular to the atomic trajectory, thus the rotation-induced phase shift is zero and, in the center-of-mass coordinate frame, it can be shown that $\phi_2^{\rm A} = -\phi_2^{\rm B}$. Similarly, at $t = 2T$, the phase is $\phi_3 = -k_{\rm eff} \theta_3 L$, where $\theta_3 = \Omega T$. Using \Eq \refeqn{Phi_tot}, the total interferometer phase shift due to the rotation is $\Phi_{\rm rot} = k_{\rm eff} (\theta_1 + \theta_3) L = -2 \, k_{\rm eff} \, \varv \, \Omega T^2$. Here, we have used the fact that the separation between Raman pulses is $L = \varv T$ with $\varv$ the initial atomic velocity at the entrance of the interferometer. A more general form of this expression, where the rotation vector $\bm{\Omega}$ is not necessarily perpendicular to the plane of the interferometer, is given by \citep{Storey-JPhysII-1994}:
\be
  \label{Phi_rot}
  \Phi_{\rm rot} = -2 (\bm{k}_{\rm eff} \times \bm{\varv}) \cdot \bm{\Omega} T^2.
\ee
Clearly, the rotation phase shift scales linearly with $\varv$ and $\Omega$, and it scales quadratically with $T$ (or $L$). This implies that the rotation sensitivity of the matter-wave interferometer scales with the enclosed area---in the same manner as an optical Sagnac interferometer. In fact, \Eq \refeqn{Phi_rot} can be recast to highlight this area dependence by defining the area vector as $\bm{A} = -(\hbar \bm{k}_{\rm eff}/M) T \times \bm{\varv}T$. Then $\Phi_{\rm rot} = 2 M \bm{A} \cdot \bm{\Omega}/\hbar$, which is equivalent to the Sagnac phase for matter waves given by \Eq \refeqn{Phi_Sagnac}.

\subsection{Space-domain or time-domain atom interferometers: atomic beams versus cold atoms}
\label{subsec:spacetime}

Following \Eq \eqref{Phi_rot}, two strategies exist for maximizing the sensitivity of the rotation sensor: increasing the atomic velocity, \ie increasing the distance $L = \varv \, T$ between the beam splitters, or increasing the interrogation time $T$. The former requires an atomic beam source and will be referred to as a space-domain interferometer. In this configuration, the Raman lasers are running continuously and the Sagnac area is defined, in practice, by physical quantities $L$ and $\varv$ (area proportional to $L^2/\varv$). The latter will instead work in the time domain and requires the use of cold atoms which can be interrogated for sufficiently long times (typically 100 ms). In this second configuration, the Raman lasers are pulsed in order to define the interaction time with the atoms, and the Sagnac area is defined by physical quantities $T$ and $\varv$ (area proportional to $\varv \, T^2$). We will give two examples of such experiments.

\paragraph{Space-domain interferometers with an atomic beam: \Refs \citep{Gustavson-ClassQuantumGrav-2000, Durfee-PRL-2006}}

By the early 2000s, Sagnac interferometers based on atomic beams had been significantly improved compared to the first experiments in the 1990s \citep{Riehle-PRL-1991, Lenef-PRL-1997, Gustavson-PRL-1997}. Specifically, the work of Gustavson \textit{et al} \citep{Gustavson-ClassQuantumGrav-2000} at Yale helped realize short-term sensitivities of $\sim 6 \times 10^{-10}$ rad/s/$\sqrt{\rm Hz}$. This gain in sensitivity arose mainly due to the implementation of a high-flux atom source. Moreover, it solved for the first time the problem of discriminating between phase shifts from rotation and from acceleration by the implementation of a counter-propagating atomic beam geometry. Since the sign of the rotation-induced phase shift given by \Eq \refeqn{Phi_rot} depends on the velocity vector, reversing the direction of the atomic beam results in a phase shift with opposite sign. Thus, by measuring the interference fringes from two separate counter-propagating sources, one can suppress via common-mode rejection parasitic phase shifts arising, for example, from the acceleration due to gravity or vibrations of the Raman laser optics.

\begin{figure}[!t]
  \centering
  \includegraphics[width=0.80\textwidth]{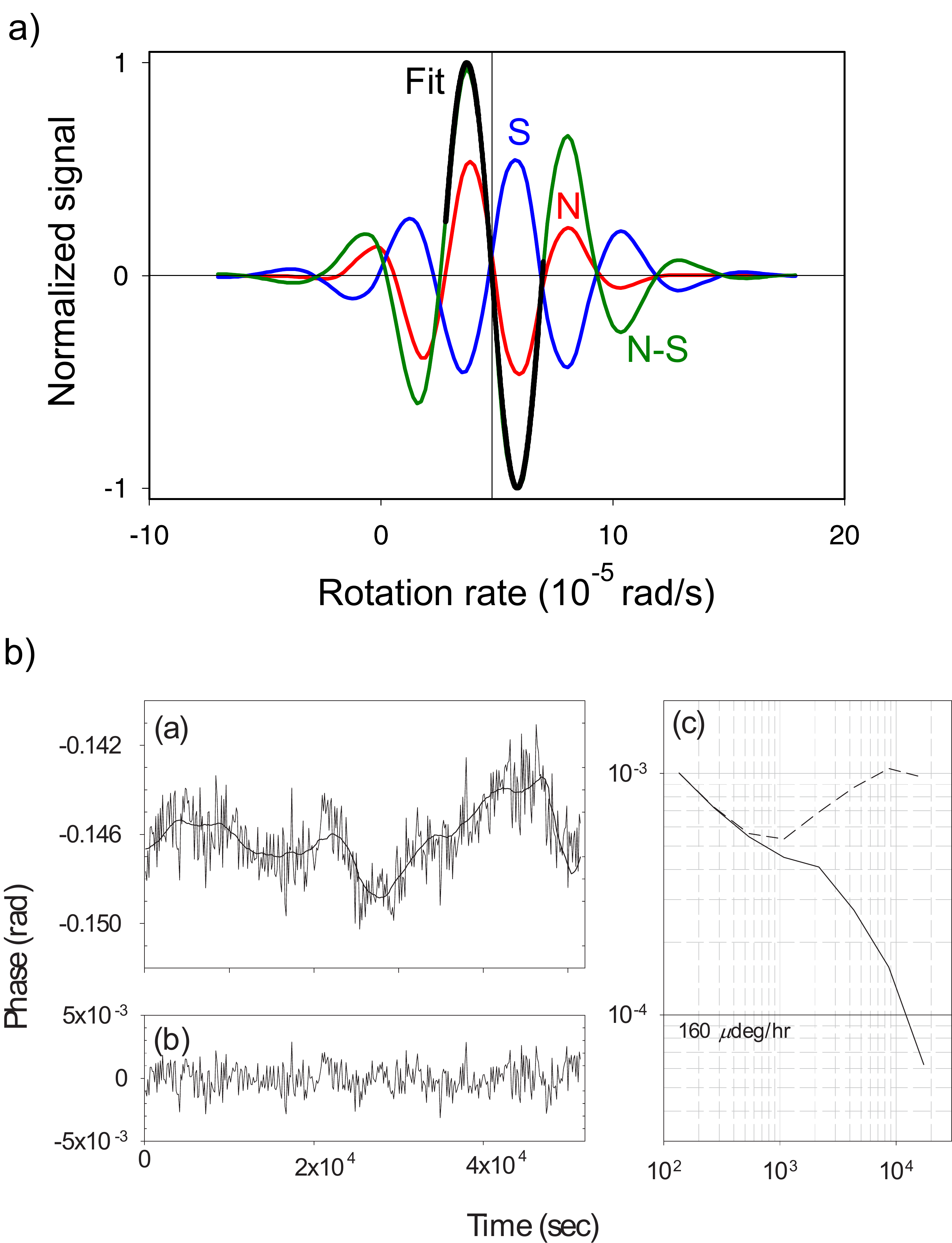}
  \caption{(Color online) Space-domain interferometer with an atomic beam. a) Interference fringes from the counter-propagating atomic beam experiments in \Ref \citep{Gustavson-ClassQuantumGrav-2000} (figure adapted from \citep{Gustavson-ClassQuantumGrav-2000}). Here, the fringes labeled ``N'' and ``S'' are from north and south facing beams, respectively, while the difference is labeled ``N-S''. A fit to this signal, shown as the solid black line, gives an estimate of the Earth's rotation rate where the line crosses zero. b) Rotation phase measurements recorded over 14 hours from \Ref \citep{Durfee-PRL-2006}. The upper plot labeled ``(a)'' shows the raw measurements compensated with $k$-reversal, along with a fit to sum of five independent temperature measurements (solid line). Plot ``(b)'' shows the temperature-compensated phase, and ``(c)'' is the Allan deviation of the rotation signal (dashed line: Allan deviation of the uncorrected data, solid line: Allan deviation of the corrected data).}
  \label{fig:Gustavson-Durfee}
\end{figure}

In an effort to further reduce systematic effects and improve the long-term accuracy of the gyroscope, an additional technique was later introduced by Durfee \textit{et al} \citep{Durfee-PRL-2006} at Stanford to eliminate spurious non-inertial phase shifts, such as those produced by magnetic fields or ac Stark effects. This involved periodically reversing the direction of $\bm{k}_{\rm eff}$ between measurements of the two interference signals from each atomic beam, which facilitated a sign reversal of the inertial phase while maintaining the sign of the non-inertial phase. Combining these four signals drastically reduced systematic shifts and long-term drift of rotation phase measurements (stability of $\sim 2.5 \times 10^{-9}$ rad/s in 15 min), at the cost of the short term sensitivity. Further correlation analysis with measured environmental variables, such as temperature, indicate that the long-term sensitivity could considerably be reduced to $\sim 3 \times 10^{-10}$ rad/s in 5 h \cite{Durfee-PRL-2006} by a correction proportional to those measurements, as shown in \Fig \ref{fig:Gustavson-Durfee}b).

\paragraph{Time-domain interferometers with laser cooled atoms: \Refs \citep{Canuel-PRL-2006,Gauguet-PRA-2009}}

\begin{figure}[!t]
  \centering
  \includegraphics[width=\textwidth]{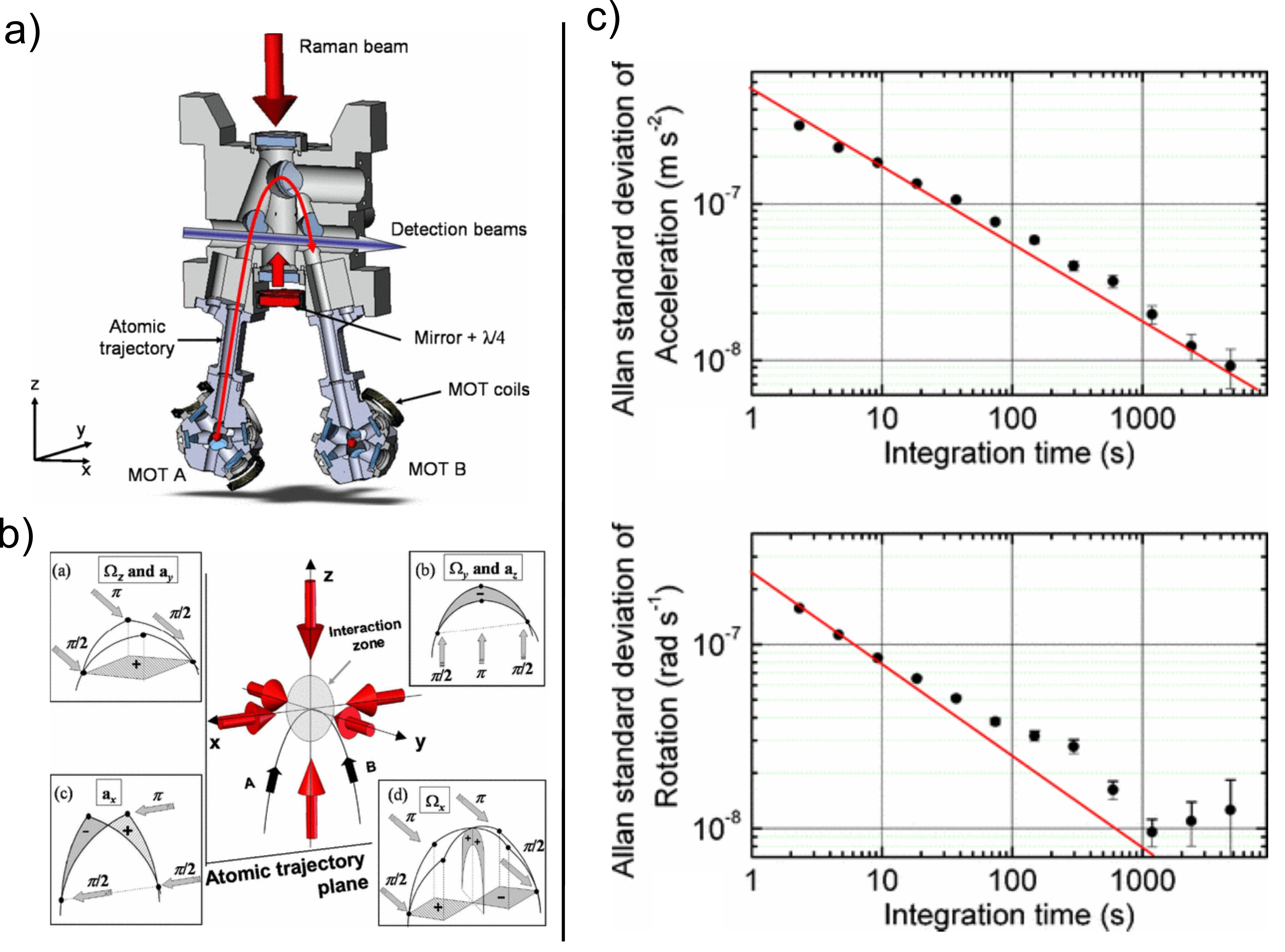}
    \caption{(Color online) a) Schematic of the SYRTE atomic gyroscope-accelerometer experiment using two cold atom clouds, from \Ref \citep{Gauguet-PRA-2009}. b) Interferometer configurations leading to information on the three axes of inertia, from \Ref \citep{Canuel-PRL-2006}. Performances of the accelerometer-gyroscope obtained in 2009 by Gauguet \textit{et al} \citep{Gauguet-PRA-2009}. The acceleration sensitivity is limited by residual vibrations of the platform (top panel), while the rotation measurement is limited by quantum projection noise (bottom panel).}
  \label{fig:SYRTEcoldatoms}
\end{figure}

In contrast to atomic gyroscopes using the propagation of atomic beams over meter-long distances, cold atom interferometers make use of the $T^2$ scaling of the gyroscope sensitivity by interrogating laser cooled atoms during $\sim 100$ ms. They allow for more compact setups and for a better control of atomic trajectories and thus of systematic effects. A pioneering experiment that started at SYRTE (France) in the early 2000s used two counter-propagating clouds of cesium atoms launched in strongly curved parabolic trajectories. Three single Raman beam pairs, pulsed in time, were successively applied in three orthogonal directions leading to the measurement of the three axes of rotation and acceleration, thereby providing a full inertial base \citep{Canuel-PRL-2006}. The SYRTE atomic gyroscope-accelerometer experiment is shown in \Fig \ref{fig:SYRTEcoldatoms}a). Figure \ref{fig:SYRTEcoldatoms}b) presents the various interrogation configurations that enable extraction of the three components of acceleration and rotation.
The short-term acceleration and rotation sensitivity of the instrument (with 1 s of integration) was first  $4.7 \times 10^{-6}$ m/s$^2$ and $2.2 \times 10^{-6}$ rad/s in the work of Canuel \textit{et al} \citep{Canuel-PRL-2006}, respectively. The setup (in particular the detection system and atom source preparation) was then improved to reach the quantum projection noise limit on the rotation measurement at the level of $2.4 \times 10^{-7}$ rad/s/$\sqrt{\rm Hz}$, and a long-term sensitivity of $1 \times 10^{-8}$ rad/s at 1000 s integration time [see \Fig \ref{fig:SYRTEcoldatoms}c), bottom panel], which was ultimately limited by the fluctuation of the atomic trajectories due to wavefront distortions of the Raman lasers \cite{Gauguet-PRA-2009}. Two other important features of this device had been tested: the linearity with the rotation rate and the independence of the rotation measurement from the acceleration. First, the evaluation of the non-linearities from a quadratic estimation of the scaling factor evolution has been demonstrated to be below $10^{-5}$. Second, the effect of the acceleration on the rotation phase shift is canceled at a level better than 76 dB when adding a well-controlled DC acceleration on the apparatus.

To conclude this section, we present in Table \ref{tab:GyroComparison} a comparison of the gyroscopes using atomic beams and cold atoms. Although the geometries are very different, the final sensitivity levels are similar (atomic beams show increased sensitivity by a factor of $\sim 3$). Furthermore, cold atoms offer better control of systematic effects and more compact setups with margins of improvements by an optimization of the geometry. In particular, an improvement of both short-term and long-term stabilities should arise from a larger average velocity, which was chosen to be very small in this first experiment (33 cm/s). In the next section, we present the new generation of cold atom experiments since 2009 aiming at improving the performances by more than one order of magnitude.

\begin{table}[!t]
  \centering
  \small
   \begin{tabular}{|l||l|l|}
    \hline
      \textbf{Domain} & Time (SYRTE, 2009) & Space (Yale, 2000 / Stanford, 2006) \\
      \hline
      \hline
      \textbf{Atomic source} & MOT & Atomic Beam \\
      \hline
      \textbf{Flux} & Low & High \\
      \hline
      \textbf{Sagnac area} & 4 mm$^2$ & 24 mm$^2$ \\
      \hline
      \textbf{Velocity} & 33 cm/s & 290 m/s \\
      \hline
      \textbf{Interferometer length} & 2.7 cm & 2 m \\
      \hline
      \textbf{Sensor size} & 0.5 m & 2.5 m \\
      \hline
      \textbf{Velocity control} & Good (molasses) & Poor (Oven) \\
      \hline
      \textbf{$T$ control}  & Very good & Poor ($T = L/\varv$) \\
      \hline
      \textbf{Acceleration sensitivity} & Very high (large $T$) & Moderate \\
      \hline
      \textbf{Acceleration rejection} & Very good ($T$ symmetric) & Moderate (asymmetry in $\varv$) \\
      \hline
      \textbf{Wavefront distortion limited} & Yes & Probably \\
      \hline
      \textbf{Short-term sensitivity (1 s)} & $2.3 \times 10^{-7}$ rad/s/Hz$^{1/2}$ & Yale (2000): $6 \times 10^{-10}$ rad/s/Hz$^{1/2}$ \\
      & & Stanford (2006): $8 \times 10^{-8}$ rad/s/Hz$^{1/2}$ \\
      \hline
      \textbf{Long-term sensitivity (15 min)} & $1.0 \times 10^{-8}$ rad/s & Yale (2000): not specified \\
      & & Stanford (2006): $2.5 \times 10^{-9}$ rad/s \\
      \hline

  \end{tabular}
   \caption{Comparison of gyroscope properties for systems based on cold atoms and atomic beams. The Yale (2000) experiment demonstrated exceptional short term sensitivity but did not demonstrate long term stability \cite{Gustavson-ClassQuantumGrav-2000}. The short term sensitivity of the Stanford (2006) experiment is extrapolated back to one second using the long term stability of $2.5 \times 10^{-9} \ \text{rad/s}$ and assuming a rotation rate stability scaling as $1/\sqrt{\tau}$ from \Ref \cite{Durfee-PRL-2006}.}
  \label{tab:GyroComparison}
\end{table}

\subsection{Latest generation of cold atom gyroscopes: \Refs \citep{LevequePhD2010, Stockton-PRL-2011, Tackmann2012, MeunierPhD2013, Tackmann2014})}
\label{subsec:newgeneration}

Following the experiments discussed previously, the strategy to enhance the sensitivity of the gyroscope essentially consists of increasing the interferometer area. Two geometries have been developed so far. First, keeping the same three Raman pulses configuration but with a straighter horizontal trajectory ($\varv = 2.8$ m/s), the gyroscope of the University of Hannover \citep{Tackmann2012} has an area 5 times larger (19 mm$^2$) with preliminary results \citep{Tackmann2014} similar to those of SYRTE.

\begin{figure}[!t]
  \centering
  \includegraphics[width=0.8\textwidth]{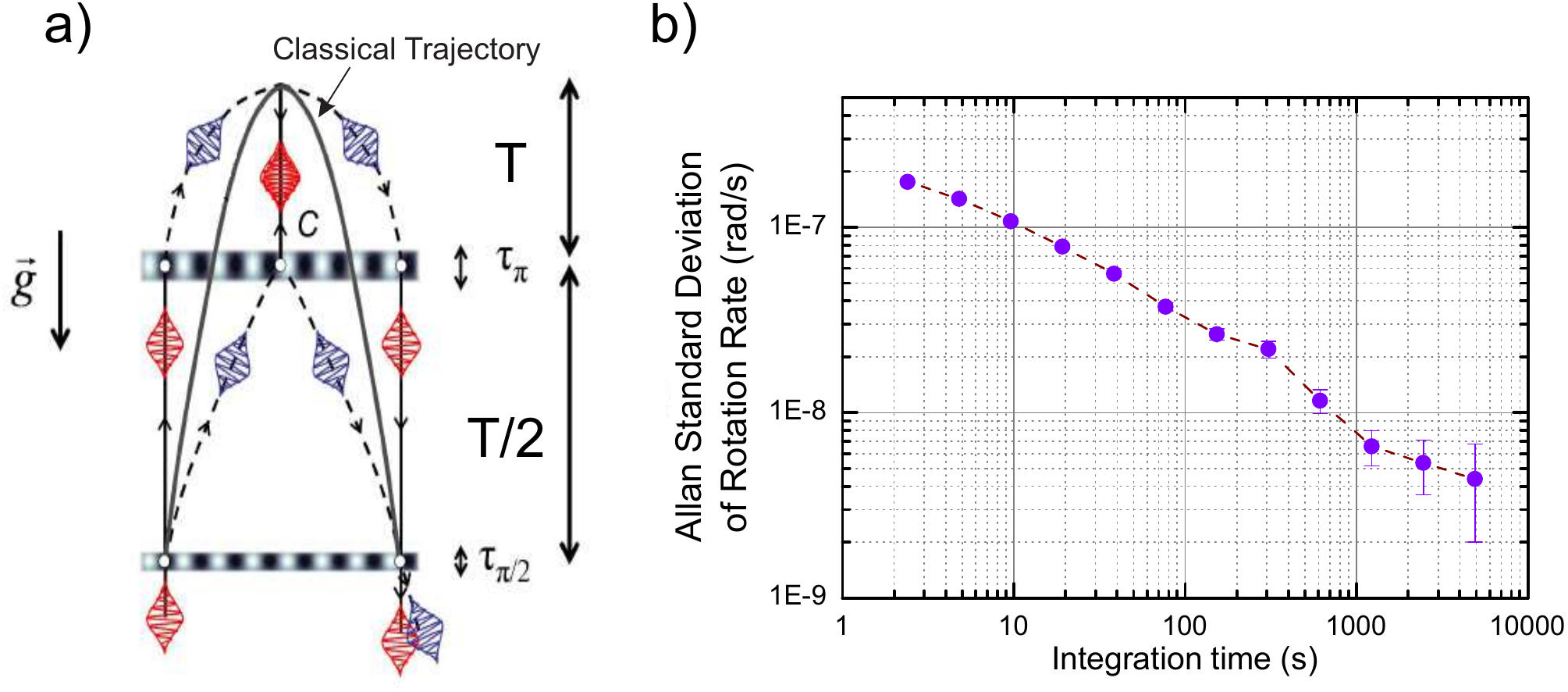}
  \caption{(Color online) a) Atomic trajectories of the SYRTE four-pulse interferometer from \Ref \citep{LevequePhD2010}. b) The Allan deviation of rotation rate measurements for an interrogation time $2T = 480$ ms. The preliminary results indicate a sensitivity of $4 \times 10^{-9}$ rad/s with 5000 s of integration time.}
  \label{fig:newgene}
\end{figure}

The second solution is based on four Raman light pulses and an atom cloud following a vertical trajectory [see \Fig \ref{fig:newgene}a)]. In that case, the atom interrogation is symmetric with respect to the apogee of the atom trajectory and is not sensitive to the DC acceleration. This new geometry was first demonstrated in \Ref \citep{Canuel-PRL-2006} and has shown improved performances in \Ref \citep{Stockton-PRL-2011}. Since the interferometer phase shift scales as $\Phi_{\rm rot} \sim k_{\rm eff} g \, \Omega \, T^3$, and the maximum possible area is 300 times larger (11 cm$^2$ with a total interrogation time of $2T = 800$ ms), substantial improvements in sensitivity are anticipated. Preliminary results presented in \Fig \ref{fig:newgene}b) have already shown a short-term sensitivity similar to the one obtained in the three-pulse configuration (improved by one order of magnitude compared to the previous four-pulse experiment), as well as an improvement in long-term stability to $4 \times 10^{-9}$ rad/s in 5000 s of integration time.

The present limit to the sensitivity arises from vibration noise of the Raman beam retro-reflecting mirrors at frequencies higher than those of the rotation signal of interest (which has a characteristic time scale of variation of several hours). The impact of the Raman mirror vibrations is a commonly encountered problem in cold atom inertial sensors and has been addressed in various works, e.g. in atomic gravimeters~\cite{Lautier2014}. The corresponding limit to the sensitivity arises from the dead time between consecutive measurements (due to the cold atom cloud preparation and detection) which results in an aliasing effect when the high frequency noise is projected onto the measurements. In other words, the dead time corresponds to a loss of information on the vibration noise spectrum, making it difficult to remove from the measurements.

\section{Conclusion and Perspectives}
\label{sec:conclusion}

After the first proof-of-principle experiments in the early 1990s, Sagnac interferometry with matter-waves has benefited from the important progress of atomic physics in the last 20 years. These advances have allowed the continuous improvement in performances of atomic gyroscopes in terms of sensitivity, long-term stability, linearity and accuracy, making atomic setups competitive or better than state-of-the-art commercial laser gyroscopes. These improvements are motivated by possible applications in inertial guidance and in geophysics. Both space- and time-domain interferometers have their own advantages. For space-domain interferometers with atomic beams this includes zero dead time between measurements, high dynamic range and a relative simplicity, versus better control of the scaling factor and smaller size for time-domain interferometers with cold atoms.

For applications in inertial navigation, the use of straight horizontal trajectories \citep{Tackmann2012} is more favorable than highly curved parabolic trajectories \citep{Gauguet-PRA-2009}. On one hand, using horizontal trajectories with fast atoms reduces the interrogation time $T$, thereby reducing the acceleration sensitivity (scaling as $T^2$) while keeping a high Sagnac scale factor (proportional to the atomic velocity), thus optimizing the ratio of rotation sensitivity over residual acceleration sensitivity. On the other hand, as demonstrated in the optical domain by laser-based gyroscopes \citep{Schreiber2011}, very-large-area atom interferometers based on highly curved parabolic trajectories \citep{LevequePhD2010,MeunierPhD2013} are of important potential interest in the field of geophysics. In this latter case, the possibility to measure rotation rates and accelerations simultaneously is advantageous in order to distinguish between fluctuations of the Earth's rotation rate and fluctuations of the projection of this rate on the measurement axis of the gyroscope. Another possibility for enhancing the Sagnac interferometer area could consist in transferring a large momentum to the atoms during the matter-wave diffraction process. Such large momentum transfer beam splitters, studied since 2008 by several groups, could result in more compact Sagnac cold atom gyroscopes of reduced interrogation times.

Nevertheless, the main limitation on increasing the sensitivity of time-domain interferometers in both applications (inertial navigation and geophysics) comes from aliasing of high-frequency noise due to measurement dead times. One solution consists of increasing the measurement repetition rate \citep{Rakholia-arXiv-2014} but at the cost of a reduction of the interrogation time and, consequently, the sensitivity. A second method could consist of hybridizing a conventional optical gyroscope with the atom interferometer in order to benefit from the large bandwidth of the former, and the long-term stability and accuracy of the latter. This method has been demonstrated in the case of the measurement of a component of acceleration by hybridizing a classical accelerometer and an atomic gravimeter \citep{Lautier2014,LautierPhD2014}. Another possibility could consist in operating a cold-atom interferometer without dead time between successive measurements in a so called joint interrogation scheme \citep{MeunierSubmitted,MeunierPhD2013}.

Besides the improvement of these Sagnac interferometers using atoms in free fall, the development of confined ultra-cold atomic sources opens the way for new types of matter-wave Sagnac interferometers in which the atoms are sustained or guided \citep{Wu-PRL-2007, Garrido2012}. Under these conditions, the interrogation time should no longer be limited by the free-fall time of the atoms in the vacuum system, and larger interferometer areas can be achieved. The present limitation on long-term stability due to wavefront distortions of the Raman laser could be lifted, since the position of the atoms will be well controlled. In contrast, the interaction with the guide or between ultra-cold atoms should bring new systematic effects which will require further study.

\section{Acknowledgements}
\label{sec:acknowledgements}

This work is supported by D\'{e}l\'{e}gation G\'{e}n\'{e}rale pour l'Armement and the French space agency CNES (Centre National d'Etudes Spatiales). B. Barrett and I. Dutta also thank CNES and FIRST-TF for financial support. The laboratory SYRTE is part of the Institut Francilien pour la Recherche sur les Atomes Froids (IFRAF) supported by R\'{e}gion Ile de France.

\bibliographystyle{elsarticle-num}
\bibliography{Sagnac_Gyro}
\end{document}